\documentclass[aps,showpacs,pra,superscriptaddress,]{revtex4}
\usepackage{amsmath}
\usepackage{amsfonts}
\usepackage{amssymb}
\usepackage{bm}
\usepackage{graphicx}

\begin{document}

\title{Periodic and solitary waves generating in optical fiber amplifiers
and fiber lasers with distributed parameters}
\author{Vladimir I. Kruglov}
\affiliation{Centre for Engineering Quantum Systems, School of Mathematics and Physics,
The University of Queensland, Brisbane, Queensland 4072, Australia}
\author{ Houria Triki}
\affiliation{Radiation Physics Laboratory, Department of Physics, Faculty of Sciences,
Badji Mokhtar University, P. O. Box 12, 23000 Annaba, Algeria}

\begin{abstract}
We study self-similar dynamics of picosecond light pulses generating in
optical fiber amplifiers and fiber lasers with distributed parameters. A
rich variety of periodic and solitary wave solutions are derived for the
governing generalized nonlinear Schr\"{o}dinger equation with varying
coefficients in the presence of gain effect. The constraint on distributed
optical fiber parameters for the existence of these wave solutions is
presented. The dynamical behaviour of those self-similar waves is discussed
in a periodic distributed amplification system. The stability of 
periodic and solitary wave solutions is also studied numerically by adding
white noise. It is proved by using the numerical split-step Fourier method
that the profile of these nonlinear self-similar waves remains unchanged
during evolution.
\end{abstract}

\maketitle

\section{Introduction}

A self-similar wave (or a similariton) is a nonlinear wave or a pulse that
maintains the same shape during propagation, even though its width and
amplitude change according to the management of system parameters \cite%
{Serkin,Zhang}. Particularly, the self-similar structure of a soliton hints
that a simple amplification is sufficient to reshape it to the original
structure instead of regenerating it \cite{Kodama}. Within this context, the
first experimental observation of beam profile reshaping due to
light-induced change of refractive index has been reported in \cite{Moll}.
Moreover, the self-similar evolution of ultrashort parabolic pulses in a
laser resonator was also experimentally observed \cite{Ilday}. In addition,
self-similarity has been demonstrated in the evolution of self-written
waveguides \cite{Kewitsch,Monro}, in the growth of Hill gratings \cite{An},
and in optical fibers \cite{Anderson}. These discoveries prove the richness
of self-similar behaviours in field of nonlinear optics.

Nowadays, studying the formation and properties of self-similar structures
has become an increasingly active field of research \cite{R1}-\cite{R6}, due
to their importance to understand widely various physical phenomena \cite%
{Barenblatt}. Notably, these chirped self-similar pulses provide a potential
application to the design of fiber optic amplifiers, optical pulse
compressors, and solitary wave-based communication links \cite{K1,D1,M1}.

A realistic description of self-similar pulse dynamics in an inhomogeneous
Kerr medium is most often based on the generalized nonlinear Schr\"{o}dinger
(NLS) equation with distributed coefficients \cite{K1,K2}. The various
parameters involved in this model which depict group velocity dispersion,
cubic nonlinearity, and gain or loss are allowed to change with the
propagation distance, thus reflecting the presence of inhomogeneity in the
real system. With the inclusion of these varying coefficients, the
underlying model becomes difficult to solve exactly. The development of
powerful methods for obtaining analytical self-similar solutions and ongoing
improvement in the analysis required to determine optical nonlinear waves
possessing physically relevant properties are thus essential for studying
the self-similar dynamics at both picosecond and femtosecond time scales.
For the purpose of identifying different nonlinear waves that may propagate
in optical fiber amplifiers and fiber lasers, the self-similarity technique
proposed by Kruglov et al. \cite{K1,K2,K3} has been widely used as a
powerful method to study the existence of self-similar solutions. We note
that thanks to this method, the determination of self-similar wave solutions
for the generalized NLS equation with distributed coefficients \cite{K1,K2}
and its many relevant variants \cite{R1,R2,Dai1,Dai2} has become possible.
For all these studies, the resulting solitary pulse solutions are shown to
exhibit special properties such as enhanced linearity in pulse frequency
chirp, self-similarity in pulse shape, and stability with respect to finite
perturbations. These self-similar wave solutions may be profitably exploited
in designing the optimal waveguiding system experiments and further
understanding of their transmission properties.

It is of interest to construct other new analytical self-similar solutions
to the generalized NLS equation with varied coefficients. No doubt, this can
help one to understand the transmission process and hence designing new
optical systems and devices. It is also very interesting to develop other
self-similarity techniques which can be applicable efficiently to search for
more nonlinear waves that can propagate self-similarly in an optical fiber
amplifier. Finding new types of self-similar propagating pulses and
constructing powerful methods able to solve the generalized NLS equation
that describes ultrashort pulse propagation phenomena in a variety of
physical situations is an interesting work. In this paper, we have developed
a new direct self-similarity method to construct a diversity of new
analytical self-similar solutions for the generalized NLS equation that
takes distributed second order dispersion, cubic nonlinearity and gain into
account. This method also allowed us to determine the self-similar variables
and formation conditions of these structures.

The structure of this paper is as follows. In Sec. II, the generalized NLS
equation modelling the evolution of picosecond optical pulses in the fiber
amplifiers and fiber lasers with distributed parameters in presence of gain
effects is presented. The set of nonlinear differential equations that
governs the dynamics of wave amplitude in the system is also derived here.
Then, in Sec. III, we present the general self-similar form of exact
periodic wave solutions and discuss the propagation properties and formation
conditions of their existence. In Sec. IV, we have obtained a set of
periodical and solitary self-similar waves governed by this generalized NLS
equation with distributed coefficients. In Sec. V, we investigate the
dynamical behaviour of the self-similar waves in a periodic distributed
amplification system for different choices of parameters. We also analyze in
Sec. VI the stability of nonlinear wave solutions by numerical simulation.
Finally, the results are summarized in Sec. VII.

\section{Scaling transformation of NLS equation}

The picosecond pulses generating in an optical amplifier with distributed
parameters are described by the following generalized NLS equation \cite%
{K1,K2,Serkin1,Wang}, 
\begin{equation}
i\frac{\partial \psi }{\partial z}=\alpha (z)\frac{\partial ^{2}\psi }{%
\partial \tau ^{2}}-\gamma (z)\left\vert \psi \right\vert ^{2}\psi +\frac{i}{%
2}g(z)\psi ,  \label{1}
\end{equation}%
where $\psi (z,\tau )$ is the complex field envelope. The variable $z$
represents the distance along direction of propagation and $\tau
=t-\int_{0}^{z}\beta _{1}(z^{\prime })dz^{\prime }$ is the time in a moving
reference frame where $\beta _{1}(z)$ is the first order distributed
dispersion. It means that we made the transformation of time $t$ to new
variable $\tau $ (retarded time) which allow to subtract the first order
partial time-derivative in the generalized NLS equation. The real
distributed parameter $\alpha (z)$ is given as $\alpha (z)=\beta _{2}(z)/2$
where $\beta _{2}(z)$ is the second order dispersion, and the parameters $%
\gamma (z)$ and $g(z)$ define the nonlinearity and gain effects. Note that
the fiber laser can be described by the same NLS equation with additional
equations defining the pumping process, feedback effect and appropriate
boundary conditions which depend on the model of laser.

It is valuable the following transformation of the wave function $\psi
(z,\tau )$ as 
\begin{equation}
\psi (z,\tau )=\chi (\zeta ,\tau ),~~~~\frac{d\zeta }{dz}=\alpha (z),
\label{2}
\end{equation}%
where $\zeta =f(z)$. Thus, the function $f(z)$ is given by 
\begin{equation}
\zeta =f(z)=\int_{0}^{z}\alpha (z^{\prime })dz^{\prime },  \label{3}
\end{equation}%
where is used the boundary condition $f(0)=0$. The generalized NLS equation (%
\ref{1}) for new wave function $\chi (\zeta ,\tau )$ has the form, 
\begin{equation}
i\frac{\partial \chi }{\partial \zeta }=\frac{\partial ^{2}\chi }{\partial
\tau ^{2}}-\tilde{\gamma}(z)\left\vert \chi \right\vert ^{2}\chi +\frac{i}{2}%
\tilde{g}(z)\chi .  \label{4}
\end{equation}%
In this NLS model equation the distributed parameters $\tilde{\gamma}(z)$
and $\tilde{g}(z)$ are given as 
\begin{equation}
\tilde{\gamma}(z)=\frac{\gamma (z)}{\alpha (z)},~~~~\tilde{g}(z)=\frac{g(z)}{%
\alpha (z)}.  \label{5}
\end{equation}%
The traveling wave solutions of the generalized NLS equation (\ref{4}) have
the form, 
\begin{equation}
\chi (\zeta ,\tau )=A(\zeta )U(\xi )\exp [i(\kappa \zeta -\delta \tau
+\theta )],  \label{6}
\end{equation}%
where $U(\xi )$ is a real amplitude function which depends on the variable $%
\xi =\tau -q\zeta $, and the parameter $q$ is connected with the inverse
velocity of the pulses in retarded frame. The parameter $q$ and modified
wave number $\kappa $ depending on the frequency shift $\delta $ are found
below. We emphasis that this (modified) wave number is introduced under the
variable $\zeta $ which is connected to the distance $z$ by Eq. (\ref{3}).
Thus, the variable $\zeta $ and modified wave number $\kappa $ have not the
standard dimensions.

It follows from Eq. (\ref{6}) that without loss of generality we can use
here the boundary condition $A(0)=1$ for the real function $A(\zeta)$. The
real parameters $\theta $ represents the phase of pulse at $\zeta=0$. The
generalized NLS equation (\ref{4}) with the wave function $\chi(\zeta,\tau )$
given in Eq. (\ref{6}) yields the ordinary differential equations for the
functions $A(\zeta)$ and $U(\xi)$ as 
\begin{equation}
\frac{1}{A(\zeta)}\frac{dA(\zeta)}{d\zeta}=\frac{1}{2}\tilde{g}(z),
\label{7}
\end{equation}%
\begin{equation}
\frac{d^{2}U}{d\xi^{2}}-\sigma U-\tilde{\gamma}(z)A^{2}(\zeta)U^{3}=0,
\label{8}
\end{equation}%
where the parameters $\sigma$ and $q$ are given by 
\begin{equation}
\sigma=\delta^{2}-\kappa,~~~~q=2\delta.  \label{9}
\end{equation}%
We note that Eq. (\ref{8}) yields the following constraint $\tilde{\gamma}%
(z)A^{2}(\zeta)=\mathrm{const}$ because the function $U$ depends on the
variable $\xi$. Thus, using the boundary condition $A(0)=1$ we can write
this constraint as 
\begin{equation}
\frac{\gamma(z)}{\alpha(z)}A^{2}(\zeta)=\frac{\gamma_{0}}{\alpha_{0}},
\label{10}
\end{equation}%
where $\gamma_{0}=\gamma(0)$, $\alpha_{0}=\alpha(0)$ and $\zeta=f(z)$.
Hence, in this case Eq. (\ref{8}) transforms to the following ordinary
nonlinear differential equation, 
\begin{equation}
\frac{d^{2}U}{d\xi^{2}}-\sigma U -\frac{\gamma_{0}}{\alpha_{0}}U^{3}=0.~~~~~~
\label{11}
\end{equation}%
We use also the following definition: 
\begin{equation}
A(f(z))\equiv B(z)~~~~ B(z)=\left(\frac{\gamma_{0}\alpha(z)}{%
\alpha_{0}\gamma(z)}\right)^{1/2}.  \label{12}
\end{equation}
Thus, Eq. (\ref{7}) can be written in the form, 
\begin{equation}
\frac{d\ln B(z)}{dz}=\frac{1}{2}g(z).  \label{13}
\end{equation}

\section{Self-similar form of wave solutions}

The equations derived in previous section lead to self-similar wave
solutions of general NLS equation (\ref{1}) with distributed coefficients.
Using Eqs. (\ref{2}), (\ref{6}) and (\ref{12}) we can write the wave
function of Eq. (\ref{1}) as 
\begin{equation}
\psi(z,\tau )=B(z)U(\xi)\exp [i(\kappa f(z)-\delta \tau +\theta)],
\label{14}
\end{equation}%
where the variables $\xi$ is 
\begin{equation}
\xi=\tau-2\delta f(z),~~~~ f(z)=\int_{0}^{z}\alpha(z^{\prime})dz^{\prime}.
\label{15}
\end{equation}%
The integration of Eq. (\ref{11}) yields the first order nonlinear
differential equations for the function $U(\xi)$ as 
\begin{equation}
\left( \frac{dU}{d\xi }\right)^{2}=\nu+\sigma U^{2}+\lambda U^{4},
\label{16}
\end{equation}
where $\nu$ is the integration constant, and the parameters $\sigma$ and $%
\lambda$ are given by 
\begin{equation}
\sigma=\delta^{2}-\kappa,~~~~\lambda=\frac{\gamma_{0}}{2\alpha_{0}}.
\label{17}
\end{equation}%
We emphasis that $\sigma$ is an arbitrary parameter here because the
modified wave number $\kappa$ is not fixed even for given frequency shift $%
\delta$.

We transform Eq. (\ref{16}) to new function $F(\xi )$ using relation as 
\begin{equation}
U^{2}(\xi )=-\frac{1}{4\lambda}F(\xi ).~~~~~~~  \label{18}
\end{equation}%
Thus, we have the nonlinear differential equation for function $F$ as 
\begin{equation}
\left( \frac{dF}{d\xi }\right) ^{2}=\rho F+\mu F^{2}-F^{3},~~~~~~~
\label{19}
\end{equation}%
where the coefficients $\rho$ and $\mu$ are 
\begin{equation}
\rho=-16\nu\lambda,~~~~\mu=4\sigma.  \label{20}
\end{equation}%
Note that in Sec. IV we find the solutions of Eqs. (\ref{16}) and (\ref{19})
for particular values of integration constant $\nu$. However, the parameter $%
\sigma$ is free under some intervals which depend on the particular solution
of Eqs. (\ref{16}) and (\ref{19}). Hence, the parameter $\mu$ is free as
well.

The integration of Eq. (\ref{13}) with the boundary condition $B(0)=1$ (see
Eq. (\ref{12})) yields 
\begin{equation}
B(z)=\exp\left(\frac{1}{2}\int_{0}^{z}g(z^{\prime})dz^{\prime}\right).
\label{21}
\end{equation}%
Hence, the nonlinearity coefficient $\gamma(z)$ is given by Eq. (\ref{12})
as 
\begin{equation}
\gamma(z)=\frac{\gamma_{0}}{\alpha_{0}}\alpha(z)\exp\left(-\int_{0}^{z}g(z^{%
\prime})dz^{\prime}\right).  \label{22}
\end{equation}%
Let us for an example define the gain as $g(z)=g_{0}=\mathrm{const}$ then
the amplitude $B(z)$ and nonlinearity parameter $\gamma(z)$ are 
\begin{equation}
B(z)=\exp\left(\frac{1}{2}g_{0}z \right),~~~~\gamma(z)=\frac{\gamma_{0}}{%
\alpha_{0}}\alpha(z)\exp\left(-g_{0}z \right).  \label{23}
\end{equation}%
One can also consider the amplitude $B(z)$ as an arbitrary function in the
class of real increasing functions. In this case the gain $g(z)$ is given by
Eq. (\ref{13}) and the nonlinearity coefficient $\gamma(z)$ follows by
constraint in Eq. (\ref{10}). Let us define the amplitude as $B(z)=1+az$
with $a>0$ then we have 
\begin{equation}
g(z)=\frac{2a}{1+az},~~~~\gamma(z)=\frac{\gamma_{0}}{\alpha_{0}}\frac{%
\alpha(z)}{(1+az)^{2}}.  \label{24}
\end{equation}%
In the following section, we show the existence of a rich set of periodical
and solitary waves governed by Eq. (\ref{1}).

\section{Periodic and solitary wave solutions}

In this section we present a set of periodic and solitary self-similar wave
solutions based on results obtained in Sec. III. The notations for variable $%
\xi$ and real amplitude $B(z)$ are presented in Eqs. (\ref{15}) and (\ref{21}%
) respectively. The modified wave number $\kappa$ and the variable $\zeta$
are given as 
\begin{equation}
\kappa=-\sigma+\delta^{2},~~~~\zeta=f(z),  \label{25}
\end{equation}%
where $\sigma$ and $\delta$ are free parameters and the function $f(z)$ is
defined in Eq. (\ref{15}). However, the frequency shift $\delta$ in
quasi-monochromatic approximation should satisfy the condition $%
\left\vert\delta\right\vert/\omega_{0}\ll 1$ where $\omega_{0}$ is the
carrier frequency.

Thus, using the above results we have obtained a set of periodical and
solitary self-similar waves governed by generalized NLS equation (\ref{1}).
We emphasis that presented periodical solutions depend on three free
parameters as the modulus $k$ of Jacobi elliptic functions and free
parameters $\sigma$ and $\delta$. The solitary wave solutions presented
below depend on two free parameters as $\sigma$ and $\delta$. However, all
these solutions depend also on two additional (trivial) free parameters as $%
\xi _{0}$ and the phase $\theta$. In these wave solutions the parameter $%
\lambda$ is given by Eq. (\ref{17}) as $\lambda=\gamma(0)/\beta_{2}(0)$.

\begin{description}
\item[\textbf{1. Periodic waves for $\protect\lambda<0$ and $\protect\sigma%
>0 $}] 
\end{description}

In the case when parameters of wave solution belong the intervals $\lambda<0$
and $\sigma>0$ we have found that Eqs. (\ref{16}) and (\ref{19}) yield the
periodic solution as 
\begin{equation}
U(\xi )=\pm [A+B\mathrm{cn^{2}}(w(\xi -\xi _{0}),k)]^{1/2},~~~~~~~
\label{26}
\end{equation}%
where the modulus $k$ of Jacobi elliptic function $\mathrm{cn}(w(\xi -\xi
_{0}),k)$ is given in the interval $0<k<1$. The free parameter (integration
constant) $\nu$ in Eqs. (\ref{16}) and (19) is given by 
\begin{equation}
\nu=\frac{\sigma^{2}(1-k^{2})}{\lambda(2-k^{2})^{2}}.  \label{27}
\end{equation}%
The parameters of periodic solution in Eq. (\ref{26}) are 
\begin{equation}
A=\frac{\sigma(k^{2}-1)}{\lambda(2-k^{2})},~~~~B=-\frac{\sigma k^{2}}{%
\lambda(2-k^{2})},~~~~~~~  \label{28}
\end{equation}%
\begin{equation}
w=\sqrt{\frac{\sigma}{2-k^{2}}}.~~~~~~~  \label{29}
\end{equation}
It follows from this solution the conditions for parameters as $\sigma>0$
and $\lambda<0$. Substitution of solution (\ref{26}) into the wave function (%
\ref{14}) yields the family of periodic wave solutions for the NLS equation (%
\ref{1}) as 
\begin{equation}
\psi (z,\tau )=\pm B(z)\lbrack A+B\mathrm{cn^{2}}(w(\xi -\xi
_{0}),k)]^{1/2}\exp [i(\kappa f(z)-\delta \tau +\theta )],  \label{30}
\end{equation}%
where $\kappa=-\sigma+\delta^{2}$ and the modulus $k$ is an arbitrary
parameter in the interval $0<k<1$ and $\xi_{0}$ is an arbitrary real
constant. We note that in the limiting cases with $k=1$ this periodic wave
reduces to a bright-type soliton solution.

\begin{description}
\item[\textbf{2. Periodic waves for $\protect\lambda<0$, $\protect\sigma<0$
and $\protect\lambda<0$, $\protect\sigma>0$}] 
\end{description}

In the case when parameters of wave solution belong the intervals $\lambda<0$%
, $\sigma<0$ and $\lambda<0$, $\sigma>0$ we have found that Eqs. (\ref{16})
and (\ref{19}) yield the periodic solution as 
\begin{equation}
U(\xi )=\pm \Lambda \mathrm{cn}(w(\xi -\xi _{0}),k),  \label{31}
\end{equation}%
where $k$ is the modulus of Jacobi elliptic function $\mathrm{cn}(w(\xi -\xi
_{0}),k)$. In this case the free parameter (integration constant) $\nu$ in
Eqs. (\ref{16}) and (19) is given by 
\begin{equation}
\nu=\frac{\sigma^{2}k^{2}(k^{2}-1)}{\lambda(2k^{2}-1)^{2}}.  \label{32}
\end{equation}%
The parameters $\Lambda $ and $w$ are given as 
\begin{equation}
\Lambda=\sqrt{\frac{-\sigma k^{2}}{\lambda(2k^{2}-1)}},~~~~w=\sqrt{\frac{%
\sigma}{2k^{2}-1}}.  \label{33}
\end{equation}%
In this solution the modulus $k$ can belong two different intervals: $0<k<1/%
\sqrt{2}$ or $1/\sqrt{2}<k<1$. It follows from this solution the conditions
for parameters as $\lambda<0$ and $\sigma<0$ when $0<k<1/\sqrt{2}$, and the
conditions for parameters are $\lambda<0$ and $\sigma>0$ when $1/\sqrt{2}%
<k<1 $. Substitution of the solution (\ref{31}) into the wave function (\ref%
{14}) yields the following family of periodic wave solutions for the
generalized NLS equation (\ref{1}):%
\begin{equation}
\psi (z,\tau )=\pm \Lambda B(z)\mathrm{cn}(w(\xi -\xi _{0}),k)\exp [i(\kappa
f(z)-\delta \tau +\theta )],  \label{34}
\end{equation}%
where $\kappa=-\sigma+\delta^{2}$ and the modulus $k$ is an arbitrary
parameter in the interval $0<k<1/\sqrt{2}$ or $1/\sqrt{2}<k<1$. In the
limiting case with $k=1$ this solution reduces to soliton solution.

\begin{description}
\item[\textbf{3. Periodic waves for $\protect\lambda>0$ and $\protect\sigma%
<0 $}] 
\end{description}

In the case when parameters of wave solution belong the intervals $\lambda>0$
and $\sigma<0$ we have found that Eqs. (\ref{16}) and (\ref{19}) yield the
periodic solution as 
\begin{equation}
U(\xi )=\pm \Lambda \mathrm{sn}(w(\xi -\xi _{0}),k),  \label{35}
\end{equation}%
where $0<k<1$. Here $\mathrm{sn}(w(\xi -\xi _{0}),k)$ is the Jacobi elliptic
function with modulus $k$. In this case the free parameter (integration
constant) $\nu$ in Eqs. (\ref{16}) and (19) is given by 
\begin{equation}
\nu=\frac{\sigma^{2}k^{2}}{\lambda(1+k^{2})^{2}}.  \label{36}
\end{equation}%
The parameters $\Lambda $ and $w$ are given by%
\begin{equation}
\Lambda=\sqrt{\frac{-\sigma k^{2}}{\lambda(1+k^{2})}},~~~~w=\sqrt{\frac{%
-\sigma}{1+k^{2}}}.  \label{37}
\end{equation}%
It follows from this solution the conditions for parameters as $\lambda>0$
and $\sigma<0 $. Substitution of the solution (\ref{35}) into the wave
function (\ref{14}) yields the following family of periodic wave solutions
for the generalized NLS equation (\ref{1}):%
\begin{equation}
\psi (z,\tau )=\pm \Lambda B(z)\mathrm{sn}(w(\xi -\xi _{0}),k)\exp [i(\kappa
f(z)-\delta \tau +\theta )],  \label{38}
\end{equation}%
where $\kappa=-\sigma+\delta^{2}$ and the modulus $k$ is an arbitrary
parameter in the interval $0<k<1$. In the limiting case with $k=1$ this
solution reduces to the kink wave solution.

\begin{description}
\item[\textbf{4. Periodic rational-elliptic waves for $\protect\lambda>0$
and $\protect\sigma<0$}] 
\end{description}

In the case when parameters of wave solution belong the intervals $\lambda>0$
and $\sigma<0$ we have found that Eqs. (\ref{16}) and (\ref{19}) yield also
the periodic rational-elliptic solution as 
\begin{equation}
U(\xi )=\pm \frac{A\mathrm{sn}\left( w(\xi -\xi _{0}),k\right) }{1+\mathrm{dn%
}\left( w(\xi -\xi _{0}),k\right) },  \label{39}
\end{equation}%
where $0<k<1$, and the parameter (integration constant) $\nu$ in this
solution is%
\begin{equation}
\nu=\frac{\sigma^{2}k^{4}}{4\lambda(2-k^{2})^{2}}.  \label{40}
\end{equation}%
The parameters $A$ and $w$ for this periodic solution are%
\begin{equation}
A=\sqrt{\frac{-\sigma k^{4}}{2\lambda(2-k^{2})}},~~~~w=\sqrt{\frac{-2\sigma}{%
2-k^{2}}},  \label{41}
\end{equation}%
where $\sigma<0$ and $\lambda>0$. Thus, Eq. (\ref{39}) yields the periodic
bounded solution of Eq. (\ref{1}) as%
\begin{equation}
\psi (z,\tau )=\pm \frac{AB(z)\mathrm{sn}(w(\xi -\xi_{0}),k)}{1+\mathrm{dn}%
(w(\xi -\xi _{0}),k)}\exp [i(\kappa f(z)-\delta \tau +\theta )],  \label{42}
\end{equation}%
where $\kappa=-\sigma+\delta^{2}$ and the modulus $k$ is an arbitrary
parameter in the interval $0<k<1$.

\begin{description}
\item[\textbf{5. Bright solitary waves for $\protect\lambda<0$ and $\protect%
\sigma>0$}] 
\end{description}

We consider here the limiting case of solution in Eq. (\ref{26}) with $k=1$.
Hence, we have the soliton solution of Eqs. (\ref{16}) and (\ref{19}) as%
\begin{equation}
U(\xi )=\pm \left( -\frac{\sigma}{\lambda}\right) ^{1/2}\mathrm{sech}(\sqrt{%
\sigma}(\xi -\xi_{0})),  \label{43}
\end{equation}%
Thus, Eq. (\ref{43}) yields the bright solitary wave solution (with $\nu=0$)
for generalized NLS equation (\ref{1}) as%
\begin{equation}
\psi (z,\tau)=\pm\left(-\frac{\sigma}{\lambda}\right)^{1/2}B(z)\mathrm{sech}(%
\sqrt{\sigma}(\xi -\xi _{0}))\exp [i(\kappa f(z)-\delta \tau +\theta )],
\label{44}
\end{equation}
where $\kappa=-\sigma+\delta^{2}$. This solitary wave exists for $\sigma>0$
and $\lambda<0$. Note that the limiting case of solution in Eq. (\ref{34})
with $k=1$ yields the wave function given in Eq. (\ref{44}) as well.

\begin{description}
\item[\textbf{6. Dark solitary waves for $\protect\lambda>0$ and $\protect%
\sigma<0$}] 
\end{description}

The limiting case with $k=1$ leads solution in Eq. (\ref{35}) to the kink
wave solution as%
\begin{equation}
U(\xi )=\pm \Lambda _{0}\mathrm{tanh}(w_{0}(\xi -\xi _{0})).  \label{45}
\end{equation}%
The parameters of this solution (with $\nu=\sigma^{2}/4\lambda$) are%
\begin{equation}
\Lambda_{0}=\left(-\frac{\sigma}{2\lambda}\right)^{1/2},~~~~w_{0}=\sqrt{%
\frac{-\sigma}{2}},  \label{46}
\end{equation}%
where $\sigma<0$ and $\lambda>0$. Hence, the kink solitary solution for Eq. (%
\ref{1}) is 
\begin{equation}
\psi (z,\tau )=\pm \Lambda _{0}B(z)\mathrm{tanh}(w_{0}(\xi -\xi _{0}))\exp
[i(\kappa f(z)-\delta \tau +\theta )],  \label{47}
\end{equation}%
where $\kappa=-\sigma+\delta^{2}$. Note that this kink solution has the form
of dark soliton for intensity $I=|\psi (z,\tau )|^{2}=\Lambda
_{0}^{2}B^{2}(z)\mathrm{tanh}^{2}(w_{0}(\xi -\xi_{0}))$.

\begin{description}
\item[\textbf{7. Dark rational-solitary waves for $\protect\lambda>0$ and $%
\protect\sigma<0$}] 
\end{description}

The limit $k\rightarrow 1$ in Eq. (\ref{39}) leads to a rational-solitary
wave of the form, 
\begin{equation}
U(\xi)=\pm \frac{A_{0}\mathrm{tanh}\left( w_{0}(\xi -\xi _{0})\right) }{1+%
\mathrm{sech}\left( w_{0}(\xi -\xi _{0})\right)}.  \label{48}
\end{equation}%
The parameters for this solitary wave (with $\nu=\sigma^{2}/4\lambda$) are 
\begin{equation}
A_{0}=\sqrt{-\frac{\sigma}{2\lambda}},\qquad w_{0}=\sqrt{-2\sigma},\qquad
\label{49}
\end{equation}%
where $\sigma<0$ and $\lambda>0$. Hence the rational-solitary wave solution
for Eq. (\ref{1}) is given by 
\begin{equation}
\psi (z,\tau )=\pm \frac{A_{0}B(z)\mathrm{tanh}\left( w_{0}(\xi -\xi
_{0})\right) }{1+\mathrm{sech}\left( w_{0}(\xi -\xi _{0})\right) }\exp
[i(\kappa f(z)-\delta \tau +\theta )],  \label{50}
\end{equation}%
where $\kappa=-\sigma+\delta^{2}$. This solitary wave has the form of dark
soliton for intensity $I=|\psi (z,\tau )|^{2}$. Remarkably, the functional
form of the solitary wave (\ref{50}) differs from the dark solitary $\text{%
tanh}$-wave.

\section{Dynamical behaviour of self-similar waves}

In this section, we discuss the dynamical behaviour of the self-similar
waves found above for a specific soliton control system. Here we take as
examples the periodic wave (\ref{30}), bright solitary wave (\ref{44}) and
dark rational-solitary wave (\ref{50}) and study the dynamical behaviour of
self-similar propagating waves through a periodically distributed nonlinear
optical fiber system for different choices of parameters. We note that
studying the pulse evolution under the influence of periodic dispersion is
important from a practical standpoint as it has application in enhancing the
signal to noise ratio and reducing Gordon-Hauss time jitter and is also
helpful in suppressing the phase matched condition for four-wave mixing \cite%
{Konar,Loomba}.

First we study the self-similar wave propagation under the influence of
distributed dispersion and constant gain. Here we consider a soliton control
system similar to that of Ref. \cite{Dai3}, where the second-order
dispersion and gain parameters are of the forms,%
\begin{equation}
\alpha (z)=\alpha _{0}\cos (pz),  \label{51}
\end{equation}%
\begin{equation}
g(z)=g_{0},  \label{52}
\end{equation}
\noindent where\ $\alpha _{0}$ and $p$ are parameters related to the
dispersion while $g_{0}$ represents the constant net gain. In this
situation, the distributed nonlinearity parameter can be determined exactly
through Eq. (\ref{22}) as

\begin{equation}
\gamma (z)=\gamma _{0}\cos (pz)\exp \left( -g_{0}z\right) ,  \label{53}
\end{equation}

Moreover, the traveling wave variable and pulse amplitude can be obtained
using Eqs. (15) and (21) as

\begin{equation}
\left. \xi =\tau -\frac{2\delta \alpha _{0}}{p}\sin (pz),\right. ~~~\left.
B(z)=\exp \left( \frac{1}{2}g_{0}z\right) .\right.  \label{54}
\end{equation}

\noindent As seen from here, the wave variable $\xi $ varies periodically
with the propagation distance $z.$ The second relation in (\ref{54}) also
shows that the amplitude $B(z)$ of self-similar propagating waves remains a
constant when the gain vanishes ($g_{0}=0$) and undergo increase when $%
g_{0}>0$.

\begin{figure}[h]
\includegraphics[width=1\textwidth]{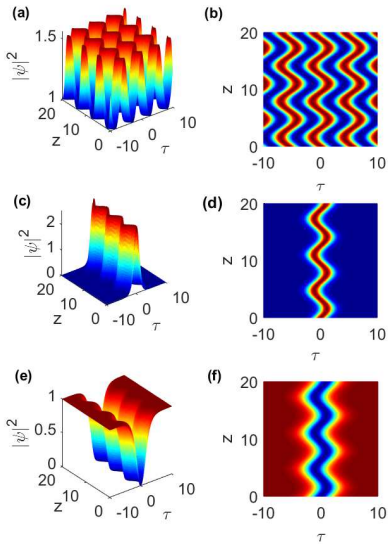}
\caption{ Evolution of (a)-(b) the periodic wave profile (\protect\ref{30}),
(c)-(d) the bright self-similar intensity wave profile (\protect\ref{44}),
and (e)-(f) the dark rational self-similar intensity wave profile (\protect
\ref{50}). The parameters are defined in the text.}
\label{FIG.1.}
\end{figure}

First, we concentrate on the most interesting situation when the fiber
system does not subject to the influence of the gain effect (i.e., $g_{0}=0$%
). Figures 1(a) and 1(b) depict the evolution of the self-similar periodic
wave (\ref{30}) for the parameter values: $\alpha _{0}=-1.2,$ $\gamma
_{0}=0.756,$ $\sigma =0.8036,$ $p=1,$ $k=0.6,$ $\delta =0.59$ and $\xi
_{0}=0.$ The results for the self-similar bright solitary wave (\ref{44})
are shown in Figs. 1(c) and 1(d) with the same parameter values as that in
Fig. 1(a) except $\delta =-0.46$. Figures 1(e) and 1(f) show the evolution
of the self-similar dark rational-solitary wave (\ref{50}) for the parameter
values: $\alpha _{0}=1.2,$ $\gamma _{0}=0.6,$ $\sigma =-0.5,$ $p=1,$ $\delta
=0.416$. These figures show clearly that the self-similar structures display
a snakelike behaviour along the propagation distance due to the presence of
periodic distributed dispersion parameter $\alpha (z).$ For such oscillatory
trajectory, the self-similar waves keep no change in propagating along
optical medium although its position oscillate periodically (which is called
\textquotedblleft Snakelike\textquotedblright\ in Ref. \cite{ZY}).

\begin{figure}[h]
\includegraphics[width=1\textwidth]{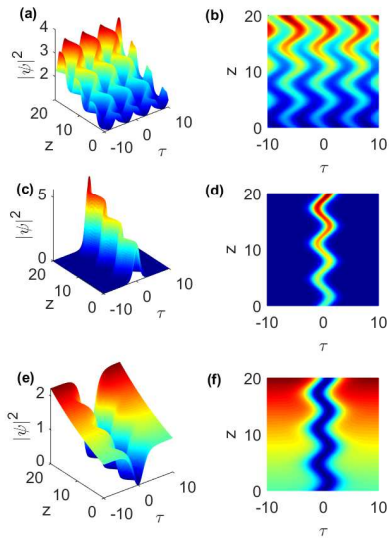}
\caption{Evolution of (a)-(b) the periodic wave profile (\protect\ref{30}),
(c)-(d) the bright self-similar intensity wave profile (\protect\ref{44}),
and (e)-(f) the dark rational self-similar intensity wave profile (\protect
\ref{50}) when $g_{0}=0.02$. The other parameters are the same as in Fig.
1(a), 1(c) and 1(e), respectively.}
\label{FIG.2.}
\end{figure}

We now consider the situation $g_{0}\neq 0$, corresponding to the presence
of gain effect in the periodic distributed system described by Eqs. (\ref{51}%
) and (\ref{52}). The evolution dynamics of the periodic wave solution (\ref%
{30}), bright solitary wave solution (\ref{44}) and dark rational-solitary
wave solution (\ref{50}) are presented in Figs. 2(a)-(b), 2(c)-(d) and
2(e)-(f), respectively for $g_{0}=0.02$. One can see that the intensity of
propagating waves increase continuously and the time shift and the group
velocity of the nonlinear waves are changing while the waves keep their
shapes in propagation along the fiber. Hence the gain parameter has no
effects on the width or shape of the nonlinear waves and affects only the
evolution of their peak.

\begin{figure}[h]
\includegraphics[width=1\textwidth]{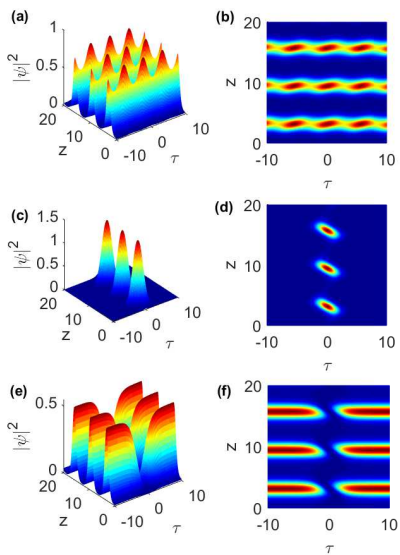}
\caption{ Evolution of (a)-(b) the periodic wave profile (\protect\ref{30}),
(c)-(d) the bright self-similar intensity wave profile (\protect\ref{44}),
and (e)-(f) the dark rational self-similar intensity wave profile (\protect
\ref{50}) when $g(z)=\sin(z)$. The other parameters are the same as in Fig.
1(a), 1(c) and 1(e), respectively.}
\label{FIG.3.}
\end{figure}

Second we investigate the self-similar periodic and solitary wave dynamics
through a distributed gain amplifier with a periodically varying gain
parameter of the form \cite{JF}:{\large \ }$g(z)=\sin (z).${\large \ }The
corresponding intensity profiles of periodic, bright and and dark
rational-solitary waves are displayed in Figs. 3(a)-(b), 3(c)-(d) and
3(e)-(f), respectively for the same values of parameters as those in Figs.1
(a), 1(b) and 1(c) respectively. For this case, we get periodic emergence of
periodic waves in the inhomogeneous fiber system due to the presence of
periodic gain, as seen in figure 3(a)-(b). It is obvious that the feature of
the bright and dark rational-solitary wave solutions is the same as shown in
Fig. 3(c)-(d) and 3(e)-(f), respectively.

We also investigated the dynamical behaviour of the self-similar periodic
and solitary waves in a distributed fiber system whose second-order
dispersion and gain parameters are distributed according to \cite{JF}:%
\begin{equation}
\alpha (z)=\tanh (z),  \label{55}
\end{equation}%
\begin{equation}
g(z)=\sin (z).  \label{56}
\end{equation}

\begin{figure}[h]
\includegraphics[width=1\textwidth]{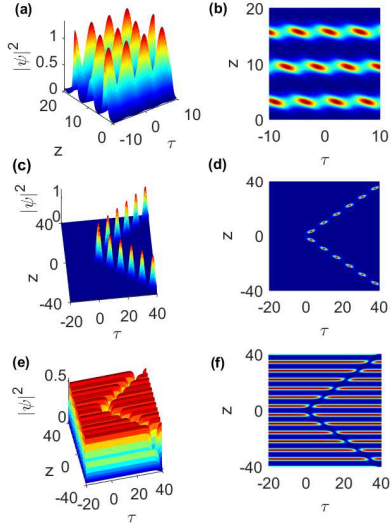}
\caption{Evolution of (a)-(b) the periodic wave profile (\protect\ref{30}),
(c)-(d) the bright self-similar intensity wave profile (\protect\ref{44}),
and (e)-(f) the dark rational self-similar intensity wave profile (\protect
\ref{50}) when $g(z)=\sin (z)$ and $\protect\alpha (z)=\tanh (z)$. The other
parameters are the same as in Fig.1(a), 1(c) and 1(e), respectively. }
\label{FIG.4.}
\end{figure}

Figures 4(a)-(b), 4(c)-(d) and 4(e)-(f) present the nonlinear evolution of
the self-similar periodic, bright- and dark-solitary-wave solutions (\ref{30}%
), (\ref{44}) and (\ref{50}) for the same values of parameters as those in
Figs.1 (a), 1(b) and 1(c) respectively. We observe that an interesting
periodic emergence of periodic, bright and dark rational-solitary waves
appear under the influence of this choice of dispersion and gain management.

It is interesting to note that we can also design some other profiles of
dispersion and gain parameters to control the dynamical behaviour of
propagating self-similar waves. We note that if the dispersion and gain
profiles are suitability chosen which may be realistic to some control
systems, we can obtain many kinds of self-similar structures with different
shapes through modulation of these parameters.

\section{Numerical stability analysis}

For the sake of completeness, we now analyze the stability of the obtained
periodic and solitary wave solutions with respect to finite perturbations.
Here we still take the periodic wave (\ref{30}), bright solitary wave (\ref%
{44}) and dark rational-solitary wave (\ref{50}) as examples to study the
structural stability of these self-similar solutions under the perturbation
of the additive white noise. Then, we perform a direct numerical simulation
of Eq. (\ref{1}) by employing the split-step Fourier method \cite{Agraw}, to
test the stability of solutions (\ref{30}), (\ref{44}) and (\ref{50}) with
initial white noise, as compared to Figs. 1(a), 1(c) and 1(e). As usual, we
put the noise onto the initial profile, then the perturbed pulse reads \cite%
{JD}: $\psi _{\text{pert}}=\psi (\tau ,0)[1+0.1\,$ random$(\tau )].$

\begin{figure}[h]
\includegraphics[width=1\textwidth]{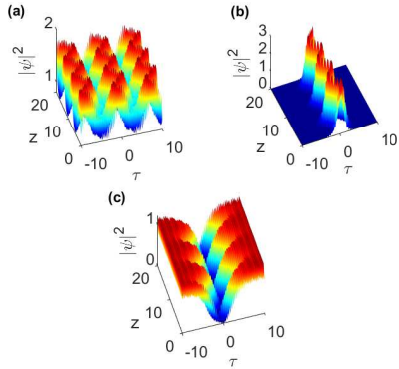}
\caption{The numerical evolution of (a) the periodic wave solution (\protect
\ref{30}), (b) the bright solitary wave solution (\protect\ref{44}), and (c)
the dark rational solitary wave solution (\protect\ref{50}) under the
perturbation of white noise whose maximal value is $0.1$. The parameters are
the same as in Figs. 1(a), 1(c) and 1(e) respectively. }
\label{FIG.5.}
\end{figure}

Figures 5(a), 5(b) and 5(c) present the evolution of nonlinear wave
solutions (\ref{30}), (\ref{44}) and (\ref{50}) under the perturbation of $%
10\%$ white noise respectively. The numerical results demonstrate that the
periodic and solitary waves can propagate stably under the initial
perturbation of white noise. Although we have shown here the results of
stability study only for three examples of the model (\ref{1}), similar
conclusions hold for other solutions as well. Therefore, we can conclude
that the solutions we obtained are stable and should be observable in
optical fiber amplifiers.

\section{Conclusion}

We have studied the self-similar dynamics of picosecond light pulses
generating in optical fiber amplifiers and fiber lasers within the framework
of the generalized nonlinear Schr\"{o}dinger equation that takes distributed
first and second order dispersions, cubic nonlinearity and gain into
account. We have developed a new self-similarity technique which enables us
to derive a rich variety of periodic and solitary wave solutions for the
model. The self-similar variables and formation conditions for the existence
of these self-similar structures are presented. The dynamical behaviour of
self-similar periodic and solitary waves has been discussed in a dispersion
periodic changing fiber for different gain profiles. The stability of these
structures is also investigated numerically by adding white noise. It is
proven that these nonlinear self-similar waves can propagate without
distortion and should therefore be observable in optical fiber amplifiers
which have a Kerr nonlinear response. In view of the increasing relevance of
the generalized NLS equation with distributed coefficients in describing the
propagation phenomena in several physical media, we anticipate that the
results presented here will open new research opportunities and may have
practical implications in experiments.

\end{document}